\documentclass[journal=jctcce,manuscript=article,layout=traditional]{achemso}
\setkeys{acs}{articletitle = true}

\usepackage{natbib}
\usepackage{amsmath}
\usepackage{amssymb}
\usepackage{graphicx}
\usepackage{dcolumn}
\usepackage{bm}
\usepackage{array}
\usepackage{color}
\usepackage{float}
\usepackage{hyperref}
\usepackage{url}
\usepackage{stfloats}

\newcolumntype{L}[1]{>{\raggedright\let\newline\\\arraybackslash\hspace{0pt}}m{#1}}
\newcolumntype{C}[1]{>{\centering\let\newline\\\arraybackslash\hspace{0pt}}m{#1}}
\newcolumntype{R}[1]{>{\raggedleft\let\newline\\\arraybackslash\hspace{0pt}}m{#1}}

\include{MyCommand}

\author{Yuan Liu}
\affiliation{Department of Chemistry, Brown University, Providence, Rhode Island 02912, USA}
\author{Minsik Cho}
\affiliation{Department of Chemistry, Brown University, Providence, Rhode Island 02912, USA}
\author{Brenda Rubenstein}
\email{brenda_rubenstein@brown.edu}
\affiliation{Department of Chemistry, Brown University, Providence, Rhode Island 02912, USA}

\title{\textit{Ab Initio} Finite Temperature Auxiliary Field Quantum Monte Carlo}

\begin{document}
\begin{abstract}
We present an \textit{ab initio} auxiliary field quantum Monte Carlo method for studying the electronic structure of molecules, solids, and model Hamiltonians at finite temperature. The algorithm marries the \textit{ab initio} phaseless auxiliary field quantum Monte Carlo algorithm known to produce high accuracy ground state energies of molecules and solids with its finite temperature variant, long used by condensed matter physicists for studying model Hamiltonian phase diagrams, to yield a phaseless, \textit{ab initio} finite temperature method. We demonstrate that the method produces internal energies within chemical accuracy of exact diagonalization results across a wide range of temperatures for H$_{2}$O (STO-3G), C$_{2}$ (STO-6G), the one-dimensional hydrogen chain (STO-6G), and the multi-orbital Hubbard model. Our method effectively controls the phase problem through importance sampling, \textit{often even without invoking the phaseless approximation}, down to temperatures at which the systems studied approach their ground states and may therefore be viewed as exact over wide temperature ranges. This technique embodies a versatile tool for studying the finite temperature phase diagrams of a plethora of systems whose properties cannot be captured by a Hubbard U term alone. Our results moreover illustrate that the severity of the phase problem for model Hamiltonians far exceeds that for many molecules at all of the temperatures studied.
\end{abstract}

\section{Introduction}
It goes without saying that most real world phenomena occur at finite temperature. Many of these phenomena, including most chemical reactions, transpire at sufficiently low temperatures that it may comfortably be assumed that the involved species' electrons remain in their ground, or at most, first few excited states. Nevertheless, the thermal distribution of electrons assumes a critical role in shaping solid phase diagrams, including the phase diagrams of many superconductors\cite{Jarrell_EPL_2001}, magnetic materials\cite{Schiffer_PRL_1995}, and trapped ultracold atoms and molecules \cite{balakrishnan2016perspective,bohn2017cold}. Thermal electrons moreover determine much of the behavior of matter under extreme conditions, such as plasmas in the warm dense matter regime often realized in inertial confinement experiments \cite{keefe1982inertial,betti2016inertial}, planetary and stellar interiors \cite{guillot1999interiors}, and materials undergoing laser irradiation \cite{glenzer2007observations}. In order to fully understand this wealth of phenomena, as well as to gauge precisely when typical ground state assumptions erode, electronic structure methods capable of not only properly accounting for temperature, but also for strong electron correlation must be developed. 

The last several decades have been marked by the emergence of a wealth of techniques adept at handling strong correlation in the ground state, including coupled cluster theory \cite{Bartlett_RevModPhys_2007}, flavors of density functional theory \cite{Perdew_AIP_2001,Medvedev_Science_2017}, and stochastic methods \cite{Foulkes_RevModPhys_2001,Motta_Review_2018}. The development of finite temperature generalizations of these methods, and in particular, generalizations capable of treating \textit{ab initio} Hamiltonians, has occurred at a far more gradual pace. Finite temperature mean field theories (MFT) \cite{mermin1963stability} and exact diagonalization (ED) techniques that account for a system's full spectrum of eigenvalues \cite{kou2014finite} have existed since the early days of quantum mechanics. Nevertheless, just as in the ground state situation, mean field theories are only reliable in the limit of weak effective electron-electron interactions and ED scales exponentially with the system size, limiting its applicability to only the smallest of quantum systems. In recent years, attempts to generalize various ground state post Hartree-Fock methods, such as many-body perturbation \cite{hirata2013kohn,santra2017finite} and coupled-cluster \cite{mandal2003finite} theories, to finite temperature  have been made, but this remains an active and growing area of research \cite{he2014finite,hermes2015finite,jaklivc1994lanczos}. Finite temperature generalizations of density functional theory are becoming increasingly popular in condensed matter physics \cite{eschrig2010t,pittalis2011exact,karasiev2014accurate,smith2015thermal,dharma2016current,cytter2018stochastic}. Nevertheless, they necessitate the development and proper benchmarking of thermal exchange correlation functionals \cite{groth2017ab,karasiev2014accurate,karasiev2016importance}. Perhaps the most resoundingly successful finite temperature algorithms have arisen from the long-standing effort to map the phase diagrams of correlated lattice models, such as the Hubbard model. Green's function methods, including dynamical mean field theory (DMFT) \cite{georges1996dynamical,kotliar2006electronic,kotliar2004strongly}, the dynamical cluster approximation (DCA) \cite{hettler1998nonlocal,hettler2000dynamical}, and most recently, self-energy embedding theory (SEET) \cite{rusakov2016self,welden2016exploring, zgid2017finite}, have driven much of our current understanding of strongly correlated material phase diagrams. These methods, however, rely upon being able to solve an impurity problem sufficiently large that it can emulate the correlations within the much larger system. In many cases, correlation lengths exceed the current size limits of common finite temperature impurity solvers \cite{Zheng_Science_2017}, reducing their overall applicability and reliability.  

Because they are naturally capable of sampling the abundance of states populated at finite temperature and can both be used as highly accurate impurity solvers as well as stand-alone techniques, quantum Monte Carlo (QMC) methods that make use of random sampling offer a path forward. Finite temperature Monte Carlo techniques, such as Path Integral Monte Carlo (PIMC) \cite{Ceperley_RevModPhys_1995} and worldline Monte Carlo \cite{Batrouni_PRB_1992}, have been resoundingly successful at mapping out the phase diagrams of bosons such as $^{4}$He because bosons lack a sign problem, the exponential decrease of signal to noise typically observed in QMC simulations of fermions that often precludes these simulations from making meaningful predictions. A generalization of real-space PIMC to fermions, restricted path integral Monte Carlo \cite{Ceperley_JStatPhys_1991,Militzer_PRL_2000}, uses constraints based upon a trial density matrix to curb the sign problem, but has been found to exhibit non-ergodic behavior. \cite{Malone_Foulkes_JCP_2015,Malone_PRL_2016} A permutation blocking PIMC algorithm that samples configuration space rather than real space has been demonstrated to circumvent these issues in plasmas.\cite{schoof2015ab,dorhheim2015pernumation} A related finite temperature variational Monte Carlo technique, VAFT, has recently been proposed, but also has not yet been generalized to \textit{ab initio} Hamiltonians. \cite{takai2016finite,claes2017finite} Some of the shortcomings of real space techniques have been overcome by recent continuous-time techniques, which now routinely provide high accuracy solutions to lattice-based impurity problems \cite{Gull_RevModPhys_2011}, but the applicability of these techniques is often limited because of the severity of the sign problem for many Hamiltonians within this framework. State-of-the-art quantum Monte Carlo methods have therefore proven themselves fully capable of sampling large finite temperature state spaces, but have yet to fully demonstrate their ability to treat \textit{ab initio} Hamiltonians.

In this paper, we develop a new generalization of Auxiliary Field Quantum Monte Carlo (AFQMC)\cite{Zhang_PRL_2003} to finite temperature \textit{ab initio} systems and benchmark its performance against a variety of chemical and model Hamiltonians that involve a range of interaction types and magnitudes\footnote{The source code can be obtained from the Brown Digital Repository \url{https://doi.org/10.7301/Z0VX0F1Z} upon request.}. Our work is motivated by the remarkable ability of the ground state phaseless AFQMC method to negotiate the sign/phase problem\footnote{In the rest of this paper, we will focus on the phaseless AFQMC method and the performance of its phaseless approximation because of the continuous Hubbard-Stratonovich transformation we employ to decouple \textit{ab initio} Hamiltonians. Nevertheless, the majority of past finite temperature auxiliary field efforts, as well as some of our concluding examples, focus upon Hamiltonians that can be decoupled via the discrete Hubbard-Stratonovich transformation, which can give rise to a sign problem. For the sake of brevity, we will view the sign problem as a specific form of the phase problem, even though its precise origin and approaches for constraining it differ.} for an increasingly large range of applications with high accuracy at a relatively low computational cost\cite{motta2017ab,motta2017towards}. In parallel, determinant quantum Monte Carlo (DQMC) has been employed for decades to illuminate finite temperature lattice physics\cite{white1988algorithm}. Applications of DQMC have overwhelmingly been limited to sign-free Hamiltonians out of fear of the growth of the sign problem \cite{Scalettar_PRB_1993} and the inaccuracies that sign-constraining approximations, such as that employed in more recently developed finite temperature AFQMC (FT-AFQMC) algorithms for the Hubbard model \cite{zhang1999finite}, may introduce. Here, we illustrate that this fear is not entirely warranted: the phase problem is not as severe as may have previously been presumed in DQMC, particularly for many \textit{ab initio} Hamiltonians, even upon cooling to the ground state. For benchmark systems ranging from water to C$_{2}$ to hydrogen chains, we find that the phase problem is so manageable that constraints are often not even required to achieve answers with milliHartree errors relative to exact results at the usual QMC $O(N^{3})-O(N^{4})$ scaling (see the Supplement for more details regarding the scaling of this method). This is yet further evidence for how controllable the phase problem for \textit{ab initio} Hamiltonians is within an overcomplete basis of nonorthogonal determinants\cite{Zhang_PRL_2003}. Even more importantly, this work establishes finite temperature phaseless AFQMC as a viable approach for studying warm dense matter and solids heretofore beyond the reach of finite temperature quantum chemical methods, or even, other stochastic approaches.

Our paper is organized as follows: In Section \ref{Methods}, we review the auxiliary field formalism for fermions at finite temperature. We then present and demonstrate how this formalism may be generalized to \textit{ab initio} Hamiltonians. In order to illustrate both the performance and versatility of our approach, in Section \ref{Results}, we benchmark our method against exact diagonalization (ED) for a plethora of first- and second-row atoms on the periodic table, as well as H$_{2}$O. We moreover present results on C$_{2}$, one-dimensional H$_{10}$ chains, and the multi-orbital Hubbard-Kanamori model, that contextualize the accuracy of our method by comparing against the results of other common alternatives. We conclude with a discussion of our findings in Section \ref{Conclusions} and leave additional supporting results and derivations for the Supplement.   

\section{Methodology \label{Methods}}
\subsection{The \emph{Ab Initio} Hamiltonian \label{Hamiltonian}}

In the following, we aim to model systems described by the \textit{ab initio} Hamiltonian
\begin{eqnarray}
\hat{H} &=& \hat{K} + \hat{V} \label{ab-initio} \\
&=& \sum_{mn} \left( T_{mn} \hat{c}_{m}^{\dagger} \hat{c}_{n} + H.C. \right) + \frac{1}{2} \sum_{mnrs} V_{mnrs} \hat{c}_{m}^{\dagger} \hat{c}_{n}^{\dagger} \hat{c}_{s} \hat{c}_{r}, 
\end{eqnarray}
where $m,n,r$, and $s$ denote spin orbital indices, $\hat{c}_{m}^{\dagger}$ creates an electron in spin orbital $m$, and $\hat{c}_{n}$ annihilates an electron in spin orbital $n$. $T_{mn}$ is the collection of all one-body integrals and $\hat{K}=\sum_{mn} \left( T_{mn} \hat{c}_{m}^{\dagger} \hat{c}_{n} + H.C. \right)$. $V_{mnrs}$ likewise denotes the collection of all two body integrals and $\hat{V}=\frac{1}{2} \sum_{mnrs} V_{mnrs} \hat{c}_{m}^{\dagger} \hat{c}_{n}^{\dagger} \hat{c}_{s} \hat{c}_{r}$. As will be used in the derivations below, the two body contributions may be re-expressed in terms of spatial orbitals and spin components

\begin{eqnarray}
\hat{V} &=& \frac{1}{2} \sum_{mnrs} V_{mnrs} \hat{c}_{m}^{\dagger} \hat{c}_{n}^{\dagger} \hat{c}_{s} \hat{c}_{r} \\
&=& \frac{1}{2} \sum_{\alpha \beta}\sum_{ijkl}^{N}V_{ijkl}^{\alpha\beta\alpha\beta}\hat{c}_{i\alpha}^{\dagger}\hat{c}_{j\beta}^{\dagger}\hat{c}_{l\beta}\hat{c}_{k\alpha}.
\end{eqnarray}
Here, $i,j,k$ and $l$ denote spatial orbital indices, $N$ denotes the number of spatial orbitals in the chosen basis, and $\alpha$ and $\beta$ denote spins. Note that, while $V_{mnrs}$ could be expanded into $V_{ijkl}^{\alpha \beta\gamma\delta}$, many of these integrals are zero in chemists' notation, leaving only the $V_{ijkl}^{\alpha\beta\alpha\beta}$ integrals behind.

\subsection{Finite Temperature AFQMC}

DQMC has long been employed to delineate the phase boundaries of fermion lattice models \cite{Scalettar_Scalapino_PRB_1986}. Recently, the DQMC algorithm has been modified into an FT-AFQMC algorithm that incorporates importance sampling and constraints on the sign problem \cite{zhang1999finite}. In the following, we review the FT-AFQMC formalism in order to ensure that our exposition is self-contained and to highlight the key modifications we have made to it in order to apply it to \textit{ab initio} Hamiltonians.  

In FT-AFQMC, it is customary\footnote{The grand canonical formalism is certainly not required. See \cite{fanto2017particle,van2006quantum} for an example of a canonical ensemble version of this approach.} to work in the grand canonical ensemble. The key quantity to sample in order to compute observables is thus the grand canonical partition function, $\Xi$, which may be expressed as
\begin{equation}
\Xi \equiv Tr(e^{-\beta(\hat{H}-\mu\hat{N})}), 
\end{equation}
where $\mu$ is the chemical potential, $\hat{N}=\sum_{i\alpha}c_{i\alpha}^{\dagger}c_{i\alpha}$, and $\beta=1/k_{B}T$ is the inverse temperature. The exponential is discretized into $L$ imaginary time pieces 
\begin{equation}
Tr(e^{-\beta(\hat{H}-\mu\hat{N})}) = Tr\Big(\lim_{\triangle\tau\to0}\prod_{l}^{L}e^{-\triangle\tau(\hat{H}-\mu\hat{N})}\Big)
\label{PartitionFunction}
\end{equation}
with $\Delta \tau = \beta/L$ so that a Suzuki-Trotter factorization of the one- and two-body contributions to the Hamiltonian may then be performed on each time slice propagator
\begin{equation}
e^{-\Delta \tau (\hat{H}-\mu \hat{N})} = e^{-\Delta \tau \hat{K}/2} e^{-\Delta \tau \hat{V}} e^{-\Delta \tau \hat{K}/2} + O(\Delta \tau^{3}). 
\label{mu_prop}
 \end{equation}
Here, we let $\hat{K}$ denote the collection of all one body operators and $\hat{V}$ denote that of all two body operators. Equation \eqref{mu_prop} may then be substituted into Equation \eqref{PartitionFunction} to yield
\begin{equation}
\Xi = Tr \left(\lim_{\Delta \tau \rightarrow 0} \prod_{l}^{L}  \left[ e^{-\Delta \tau \hat{K}/2} e^{-\Delta \tau \hat{V}} e^{-\Delta \tau \hat{K}/2} \right] \right). 
\label{Partition_Function}
\end{equation}
As the imaginary time step $\triangle\tau \to 0$, the exact partition function is recovered. 

The key to simplifying this expression used by DQMC and all of its related modern-day extensions is the Hubbard-Stratonovich (HS) Transformation. The HS Transformation enables an exponential of a two body operator to be re-expressed as either a sum (if the transformation is discrete) or integral (if the transformation is continuous) of one body operators that are functions of so-called auxiliary fields \cite{Hirsch_PRB_1983}. In conventional DQMC, as is commonly applied to Hubbard Hamiltonians that contain only density-density interactions, a discrete version of this transformation is most often employed because of its relative efficiency \cite{Buendia_PRB_1986}. Where our algorithm differs from these implementations is in its use of a continuous HS Transformation.    

\subsection{The Continuous Hubbard-Stratonovich Transformation for Decoupling \textit{Ab Initio} Hamiltonians}

In order to simplify the partition function into a form amenable to sampling for an $\textit{ab initio}$ Hamiltonian, a continuous HS Transformation must instead be performed. This is because general two body operators are not amenable to discrete transformations. The continuous HS transformation may be written as
\begin{equation}
e^{-\Delta \tau \lambda \hat{v}^{2}/2} = \frac{1}{\sqrt{2\pi}}\int_{-\infty}^{\infty} d\phi e^{-\phi^{2}/2} e^{\phi \sqrt{-\lambda \Delta \tau} \hat{v}}, 
\label{HS}
\end{equation}
where $\lambda$ denotes a constant, $\hat{v}$ denotes a one body operator, and $\phi$ denotes a Gaussian-distributed auxiliary field. This transformation implies that, as long as a two body interaction can be rewritten as the square of one body operators, it can be decoupled.  While not inherently obvious, it can be shown that \textit{ab initio} potentials may be decoupled into such a sum of squares of one body operators \cite{Motta_Review_2018}. This may be done by first reordering the potential operators so that creation and annihilation operators are paired 
\begin{align}
\hat{V} =& \frac{1}{2}\sum_{\alpha\beta}\sum_{ijkl}^{N}V_{ijkl}^{\alpha\beta\alpha\beta}\hat{c}_{i\alpha}^{\dagger}\hat{c}_{j\beta}^{\dagger}\hat{c}_{l\beta}\hat{c}_{k\alpha} \nonumber \\ 
=& -\frac{1}{2}\sum_{\alpha\beta}\sum_{ijkl}^{N}V_{ijkl}^{\alpha\beta\alpha\beta}\hat{c}_{i\alpha}^{\dagger}\hat{c}_{j\beta}^{\dagger}\hat{c}_{k\alpha}\hat{c}_{l\beta} \nonumber \\
=& \frac{1}{2}\sum_{\alpha\beta}\sum_{ijkl}^{N}V_{ijkl}^{\alpha\beta\alpha\beta}\hat{c}_{i\alpha}^{\dagger}\hat{c}_{k\alpha}\hat{c}_{j\beta}^{\dagger}\hat{c}_{l\beta} \nonumber  \\
&- \frac{1}{2}\sum_{\alpha}\sum_{ij}^{N}\big(\sum_{k}^{N}V_{ikkj}^{\alpha\alpha\alpha\alpha}\big)\hat{c}_{i\alpha}^{\dagger}\hat{c}_{j\alpha} \nonumber \\
=& \frac{1}{2}\sum_{\alpha\beta}\sum_{ijkl}^{N}V_{ijkl}^{\alpha\beta\alpha\beta}\hat{c}_{i\alpha}^{\dagger}\hat{c}_{k\alpha}\hat{c}_{j\beta}^{\dagger}\hat{c}_{l\beta} -  \sum_{\alpha} \hat{\rho}_{0}^{\alpha}
\label{eqv}
\end{align}
in which $\hat{\rho}_{0}^{\alpha} =\frac{1}{2}  \sum_{ij}^{N}(\sum_{k}^{N}V_{ikkj}^{\alpha\alpha\alpha\alpha})c_{i\alpha}^{\dagger}c_{j\alpha}$. Since the $\hat{\rho}_{0}^{\alpha}$s are entirely comprised of one body operators, they can be combined into $\hat{K}$. The two body terms may be decoupled by recasting the $V_{ijkl}^{\alpha \beta \alpha \beta}$ into a Hermitian supermatrix of dimension $(2N)^{2}\times(2N)^{2}$ with two sets of indices, $V_{(i\alpha, k\alpha), (l\beta, j\beta)}$. This supermatrix may then be decomposed via exact diagonalization (or an alternative decomposition method) into the form
\begin{align}
V_{(i\alpha,k\alpha),(l\beta,j\beta)} = \sum_{\gamma}^{(2N)^2} R_{(i\alpha,k\alpha)\gamma} \lambda_{\gamma} R_{(l\beta,j\beta)\gamma}^{*},
\label{Diagonalization}
\end{align}
where $\lambda_{\gamma}$ is the $\gamma$-th eigenvalue, and $R_{(i\alpha,k\alpha)\gamma}$ is the $(i\alpha,k\alpha)$-th element of the $\gamma$-th eigenvector. Because the supermatrix is semi-positive definite for \textit{ab initio} Hamiltonians (see the Supplement for a proof), $\lambda_{\gamma} \geq 0$. This can be reinserted into Equation \eqref{eqv} to obtain
\begin{align}
\hat{V} = \frac{1}{2} \sum_{\gamma}^{(2N)^{2}} \left[ \sum_{ik\alpha} R_{(i\alpha, k\alpha)\gamma} \hat{c}_{i\alpha}^{\dagger} \hat{c}_{k\alpha} \right]  \lambda_{\gamma} \left[ \sum_{lj\beta} R^{*}_{(l\beta, j\beta)\gamma} \hat{c}_{j\beta}^{\dagger} \hat{c}_{l\beta}   \right] -  \sum_{\alpha} \hat{\rho}_{0}^{\alpha}.
\end{align}
Let $\hat{\rho}_{\gamma} \equiv \sum_{ik\alpha} R_{(i\alpha,k\alpha)\gamma} \hat{c}_{i\alpha}^{\dagger} \hat{c}_{k\alpha}$. Then, $\hat{\rho}^{\dagger}_{\gamma} \equiv \sum_{jl\beta} R_{(l\beta, j\beta)\gamma}^{*} \hat{c}_{j\beta}^{\dagger} \hat{c}_{l\beta}$ and 
\begin{equation}
\hat{V} = \frac{1}{4} \sum_{\gamma}^{(2N)^{2}} \lambda_{\gamma} \{ \hat{\rho}_{\gamma}, \hat{\rho}_{\gamma}^{\dagger} \}-  \sum_{\alpha} \hat{\rho}_{0}^{\alpha}. 
\end{equation}
Manipulating this (see the Supplement for more details) yields the desired quadratic expression for the two body operator
\begin{equation}
\hat{V} = \frac{1}{8}\sum_{\gamma}^{(2N)^2}\lambda_{\gamma}[(\hat{\rho}_{\gamma}+\hat{\rho}_{\gamma}^{\dagger})^2-(\hat{\rho}_{\gamma}-\hat{\rho}_{\gamma}^{\dagger})^2] - \sum_{\alpha} \hat{\rho}_{0}^{\alpha}. 
\label{recast}
\end{equation}
This form is now directly amenable to the continuous HS Transformation of Equation \eqref{HS}. This yields
\begin{align}
e^{-\frac{1}{8}\sum_{\gamma}^{(2N)^2}\lambda_{\gamma} \Delta \tau [(\hat{\rho}_{\gamma}+\hat{\rho}_{\gamma}^{\dagger})^2-(\hat{\rho}_{\gamma}-\hat{\rho}_{\gamma}^{\dagger})^2]} =\prod_{\gamma}^{(2N)^2} & \big(\frac{1}{\sqrt{2\pi}}\big)^2\iint_{-\infty}^{\infty}  d\phi_{\gamma-} d\phi_{\gamma+} e^{-\frac{\phi_{\gamma+}^2+\phi_{\gamma-}^2}{2}} 
\nonumber \\
&\times e^{i\frac{\sqrt{\triangle\tau\lambda_{\gamma}}}{2}\phi_{\gamma+}(\hat{\rho}_{\gamma}+\hat{\rho}_{\gamma}^{\dagger})} 
e^{\frac{\sqrt{\triangle\tau\lambda_{\gamma}}}{2}\phi_{\gamma-}(\hat{\rho}_{\gamma}-\hat{\rho}_{\gamma}^{\dagger})}.
\label{HS-trans}
\end{align}
As two sets of exponentials must be decoupled, two sets of HS fields, $\phi_{\gamma-}$ and $\phi_{\gamma+}$, are present in the above. If $\vec{\phi}\equiv \{\phi_{\gamma+},\phi_{\gamma-}\}$, where $\gamma=1,2,...,(2N)^2$ is defined as the full set of auxiliary fields at a given imaginary time step, then all of the one body operators and Gaussians may be combined to yield
\begin{align}
e^{-\frac{1}{8}\sum_{\gamma}^{(2N)^2}\lambda_{\gamma} \Delta \tau [(\hat{\rho}_{\gamma}+\hat{\rho}_{\gamma}^{\dagger})^2-(\hat{\rho}_{\gamma}-\hat{\rho}_{\gamma}^{\dagger})^2]} = \int_{-\infty}^{\infty} d \vec{\phi} p(\vec{\phi}) \hat{B}(\vec{\phi}),  
\label{Final_Potential_Exponential}
\end{align}
where
\begin{equation}
\hat{B}(\vec{\phi}) = e^{i\frac{\sqrt{\triangle\tau\lambda_{\gamma}}}{2}\phi_{\gamma+}(\hat{\rho}_{\gamma}+\hat{\rho}_{\gamma}^{\dagger})}e^{\frac{\sqrt{\triangle\tau\lambda_{\gamma}}}{2}\phi_{\gamma-}(\hat{\rho}_{\gamma}-\hat{\rho}_{\gamma}^{\dagger})}
\label{Operator_Equation}
\end{equation}
and
\begin{equation}
p(\vec{\phi}) = e^{-\frac{\phi_{\gamma+}^2+\phi_{\gamma-}^2}{2}}.
\label{prob}
\end{equation}
Note that the up and down contributions to $\hat{B}(\vec{\phi})$ may further be partitioned into a product of up and down one body pieces $\hat{B}_{\uparrow}(\vec{\phi}) \hat{B}_{\downarrow}(\vec{\phi})$. Equation \eqref{Final_Potential_Exponential} may next be substituted into Equation \eqref{Partition_Function} to yield
\begin{equation}
\Xi \approx Tr\Big[\prod_{l}^{L}\int_{-\infty}^{\infty}d\vec{\phi}_{l} p(\vec{\phi}_{l})\hat{B}_{\uparrow}(\vec{\phi}_{l}) \hat{B}_{\downarrow}(\vec{\phi}_{l})\Big].  
\label{trace}
\end{equation}
Taking the trace over fermion operators \cite{Hirsch_PRB_1985} produces the final expression for the partition function 
\begin{align}
\Xi &\approx \int_{-\infty}^{\infty} d\vec{\phi}_{l} p(\vec{\phi}_{l}) Det[I + B_{\uparrow}(\vec{\phi}_{L})...B_{\uparrow}(\vec{\phi}_{2}) B_{\uparrow}(\vec{\phi}_{1})] \nonumber \\
& \times Det[I + B_{\downarrow}(\vec{\phi}_{L}) ...B_{\downarrow}(\vec{\phi}_{2}) B_{\downarrow}(\vec{\phi}_{1})]. 
\end{align}

\subsection{Monte Carlo Sampling}

This resulting grand canonical partition function is a highly multidimensional integral over auxiliary field space. The most efficient way to obtain observables such as energies and correlation functions from the partition function is thus to employ Monte Carlo sampling. While it is most common in conventional DQMC to sample all fields at once, we follow a more recent FT-AFQMC algorithm \cite{zhang1999finite} and sample the fields in a step by step and orbital by orbital fashion. This affords us the option of applying a constraint, such as the phaseless constraint described below, at each interval. In particular, each walker (e.g., random sample) in the simulations is initialized to have a weight of 1 and a trial density matrix, constructed from $L$ short-time propagators, $\hat{B}_{T}$, such that the initial determinants may be written as $Det[I + B_{T}...B_{T} B_{T}]$. Note that, throughout this work, we use a mean field trial density matrix. At each time slice and orbital, a new auxiliary field is sampled and the corresponding trial density matrix is replaced with an updated one body operator. Let $\phi_{ik}$ denote all of the fields sampled at time slice $k$ for orbital $i$, and $M_{ik}^{\alpha}$ denote the resulting determinant 
\begin{equation}
M^{\alpha}_{ik}=Det\Big[I+\Big(\prod_{l=1}^{L-k}B_{\alpha}^{T}\Big)B_{\alpha}(\phi_{ik}...\phi_{1k})...B_{\alpha}(\vec{\phi}_{2})B_{\alpha}(\vec{\phi}_{1})\Big]. 
\end{equation}
$\alpha$ denotes spin, as before. As each field is sampled, the walker weight is multiplied by a factor, $W(\phi_{ik})$, the ratio of the newly updated determinants to the previous determinants
\begin{equation}
W(\phi_{ik})=\frac{M^{\uparrow}_{ik}M^{\downarrow}_{ik}}{M^{\uparrow}_{(i-1)k}M^{\downarrow}_{(i-1)k}}.
\label{weight}
\end{equation}
Once all fields are sampled, each walker's observables may be computed based upon its final determinants. A weighted average may then be obtained over all walker determinants. This process, starting from the trial density matrices, may then be repeated until observables of interest are converged. 

\subsection{Background Subtraction}
As is clear from Equation \eqref{HS}, the positive eigenvalues obtained from a continuous HS Transformation of a fully \textit{ab initio} Hamiltonian result in complex one body propagators. This produces walkers with complex weights that span the complex plane. Complex weights may cancel each other when averaged, resulting in noise that is large and even uncontrollable. These are the symptoms of the so-called phase problem (a generalization of the sign problem for real-valued Monte Carlo algorithms to the complex plane). While the phase problem is insurmountable, it may be mitigated through background subtraction and importance sampling \cite{Purwanto_PRE_2004,Motta_Review_2018}.

In background subtraction, the magnitude of the imaginary part of the propagators is substantially reduced by subtracting a background estimate of the true densities. In particular, subtracting $\langle \hat{\rho}_{\gamma} + \hat{\rho}_{\gamma}^{\dagger} \rangle_{MF}$ and $\langle \hat{\rho}_{\gamma} - \hat{\rho}_{\gamma}^{\dagger} \rangle_{MF}$ from $\hat{\rho}_{\gamma} + \hat{\rho}_{\gamma}^{\dagger}$ and $\hat{\rho}_{\gamma} - \hat{\rho}_{\gamma}^{\dagger}$ in Equation \eqref{recast}, leads to a new expression for the two body term
\begin{eqnarray}
\hat{H}_{2}&=&\frac{1}{8}\sum_{\gamma}^{(2N)^2}\lambda_{\gamma}\bigg\{\big[(\hat{\rho}_{\gamma}+\hat{\rho}_{\gamma}^{\dagger})-\langle\hat{\rho}_{\gamma}+\hat{\rho}_{\gamma}^{\dagger}\rangle_{MF}\big]^{2} - \big[(\hat{\rho}_{\gamma}-\hat{\rho}_{\gamma}^{\dagger})-\langle\hat{\rho}_{\gamma}-\hat{\rho}_{\gamma}^{\dagger}\rangle_{MF}\big]^{2}\bigg\},
\label{new-tbd} 
\end{eqnarray}
where the $\langle \cdot \rangle_{MF}$ expectation values are taken with respect to the mean field trial density matrices.

As detailed in the Supplement, recasting H$_{2}$ in this form, engenders new one body terms that may be combined with the first term in Equation \eqref{ab-initio}. The continuous HS transform can then be applied to this new two body term to decouple it into one body operators, and calculate the partition function and physical observables.

The $\langle \hat{\rho}_{\gamma} + \hat{\rho}_{\gamma}^{\dagger} \rangle_{MF}$ and $\langle \hat{\rho}_{\gamma} - \hat{\rho}_{\gamma}^{\dagger} \rangle_{MF}$ used in the background subtraction can, in principle, be arbitrary. Nevertheless, the closer the trial densities are to the exact densities, the smaller the fluctuations in the exponential and the better the results will be. As will be demonstrated in the following section, mean field trial densities are sufficient for suppressing most of the fluctuations. However, our results may be systematically improved by employing more accurate trial density matrices.

\subsection{Importance Sampling and the Phaseless Approximation}

For strongly correlated systems at low temperatures, background subtraction may be not enough to control the phase problem effectively. In such cases, importance sampling may also need to be employed. In the context of AFQMC, importance sampling is used to dynamically adjust the center of the Gaussian distribution from which the auxiliary fields are sampled according to the current estimate of the wave function so as to sample the ``most important'' auxiliary fields at each interval. The constant shift introduced for importance sampling is called the force bias \cite{Purwanto_PRE_2004}.  

The force bias shifts, $\bar{\phi}_{\gamma +}$ and $\bar{\phi}_{\gamma -}$, may be incorporated into the propagator given by Equation  \eqref{HS-trans} by subtracting them from the Gaussian-distributed auxiliary fields at each time slice and for each $\gamma$
\begin{align}
& \hat{B}(\phi_{\gamma+}-\bar{\phi}_{\gamma+},\phi_{\gamma-}-\bar{\phi}_{\gamma-}) \nonumber \\ 
=& e^{i\frac{\sqrt{\triangle\tau\lambda_{\gamma}}}{2}(\phi_{\gamma+}-\bar{\phi}_{\gamma+}) \big[ \hat{\rho}_{\gamma}+\hat{\rho}_{\gamma}^{\dagger} - \langle \hat{\rho}_{\gamma}^{\dagger} + \hat{\rho}_{\gamma} \rangle \big]} e^{\frac{\sqrt{\triangle\tau\lambda_{\gamma}}}{2}(\phi_{\gamma-}-\bar{\phi}_{\gamma-}) \big[\hat{\rho}_{\gamma}-\hat{\rho}_{\gamma}^{\dagger} - \langle \hat{\rho}_{\gamma} - \hat{\rho}_{\gamma}^{\dagger} \rangle \big]} 
\end{align}
Inserting background subtraction and force bias terms into Equation \eqref{Operator_Equation} results in the final expression for the two body exponential used throughout this work 
\begin{align}
e^{-\triangle\tau\hat{H}_{2}}
=\prod_{\gamma}^{(2N)^2}\frac{1}{2\pi}\int\int_{-\infty}^{\infty}d\phi_{\gamma-} d\phi_{\gamma+}p(\phi_{\gamma+},\phi_{\gamma-})W'(\phi_{\gamma+},\bar{\phi}_{\gamma+},\phi_{\gamma-},\bar{\phi}_{\gamma-}) \hat{B}(\phi_{\gamma+}-\bar{\phi}_{\gamma+},\phi_{\gamma-}-\bar{\phi}_{\gamma-}). 
\label{importancesampling}
\end{align}
In the above, $W'(\phi_{\gamma+},\bar{\phi}_{\gamma+},\phi_{\gamma-},\bar{\phi}_{\gamma-})$ denotes an additional weight factor which contains an amalgam of all shift-related and background subtraction constants (see the Supplement for more details) and $p(\phi_{\gamma+}, \phi_{\gamma-})$ is the same Gaussian distribution as used in Equation \eqref{prob}. When performing Metropolis MC, it is the total weight, which is the product of the weight given in Equation \eqref{weight} and $W'$, that is ultimately sampled. It can be approximately shown that optimal importance sampling is achieved when the shift is set equal to the expectation value of the corresponding one body operator, $\bar{\phi}_{i} \approx -\langle \hat{v}_{i} \rangle$. The optimal field shift for each HS transform in Equation \eqref{importancesampling} is then
\begin{align}
\bar{\phi}_{\gamma +}& \approx -\Big \langle i\frac{\sqrt{\triangle\tau\lambda_{\gamma}}}{2}\big[(\hat{\rho}_{\gamma} + \hat{\rho}_{\gamma}^{\dagger})-\langle \hat{\rho}_{\gamma} + \hat{\rho}_{\gamma}^{\dagger} \rangle_{MF} \big] \Big\rangle,  \nonumber \\
\bar{\phi}_{\gamma-}& \approx -\Big\langle \frac{\sqrt{\triangle\tau\lambda_{\gamma}}}{2}\big[(\hat{\rho}_{\gamma}-\hat{\rho}_{\gamma}^{\dagger})-\langle \hat{\rho}_{\gamma}-\hat{\rho}_{\gamma}^{\dagger} \rangle_{MF} \big] \Big\rangle .
\end{align}
The overall expectation value is typically computed using the density matrices calculated during a previous propagation step, while the $\langle \cdot \rangle_{MF}$ expectation values are taken with respect to the mean field trial density matrices, as described above. As illustrated in the following, importance sampling dramatically reduces the statistical error bars we would otherwise observe. 

Even making use of these advanced numerical techniques, phase problems still typically emerge during simulations of strongly correlated systems at sufficiently low temperatures\cite{rubenstein2012finite,Zhang_PRL_2003}. To control the phase problem, the phaseless approximation may be invoked\cite{Zhang_PRL_2003}. Similar in spirit to the constrained path approximation used to curb the sign problem, the phaseless approximation introduces a constraint that confines all walkers to the positive real axis. This is accomplished by first calculating the total weight (from Equation \eqref{weight} and $W'$) after each propagation step for each walker, which is typically complex, and then projecting it onto the real axis. Moreover, as in the constrained path approximation, any walkers remaining with negative weights after this projection are killed, leaving only walkers with positive, real weights behind. Because our propagators may also become complex, applying the phaseless approximation to the weights alone does not fully resolve the phase problem. To guarantee that our observables are physical, we furthermore ignore their typically small complex components during measurements.

As further described below, one of the key findings of this work is that, for many common molecules and models even at temperatures that near the ground state, we do \textit{not} need to employ the phaseless approximation to obtain meaningful results. Consequently, the majority of the results reported below are produced using background subtraction and importance sampling alone.    
 
As in previous finite temperature DQMC algorithms, we also employ a birth/death population control scheme to control walker weights and stabilize our product of propagators \cite{white1989numerical}. 

\subsection{Observables}
In order to calculate energies, the expectation value of the Hamiltonian must be taken. Drawing upon Wick's Theorem to simplify expectation values over products of four or more fermion operators, the energy may be expressed as
\begin{equation}
\big\langle\hat{H}\big\rangle=\sum_{\alpha}\sum_{ij}^{N}T_{i\alpha,j\alpha}D_{ij}^{\alpha}+\frac{1}{2}\sum_{\alpha\ne\beta}\sum_{ijkl}^{N}V_{ijkl}^{\alpha\beta\alpha\beta}D_{ik}^{\alpha}D_{jl}^{\beta}+\frac{1}{2}\sum_{\alpha}\sum_{ijkl}^{N}V_{ijkl}^{\alpha\alpha\alpha\alpha}(D_{ik}^{\alpha}D_{jl}^{\alpha}-D_{il}^{\alpha}D_{jk}^{\alpha}). 
\end{equation}
$D_{ij}^{\alpha}$ denotes the density matrix of electrons with spin $\alpha$ in the above. These density matrices may be obtained by noting that they are directly related to the equal time, one electron Green's functions, 
$D_{ij}^{\alpha} = \delta_{ij} - G_{ji}^{\alpha}$, where
\begin{align}
G_{ij}^{\alpha}&=\frac{Tr[\hat{c}_{i\alpha}\hat{c}_{j\alpha}^{\dagger}\hat{B}_{\alpha}(\vec{\phi}_{L})\hat{B}_{\alpha}(\vec{\phi}_{L-1})...\hat{B}_{\alpha}(\vec{\phi}_{1})]}{Tr[\hat{B}_{\alpha}(\vec{\phi}_{L})\hat{B}_{\alpha}(\vec{\phi}_{L-1})...\hat{B}_{\alpha}(\vec{\phi}_{1})]} \nonumber \\
&=\Big[\frac{I}{I+B_{\alpha}(\vec{\phi}_{L})B_{\alpha}(\vec{\phi}_{L-1})...B_{\alpha}(\vec{\phi}_{1})}\Big]_{ij} .
\label{GF}
\end{align}
The occupancies of electrons with spin $\alpha$ reported throughout the paper are obtained by summing the diagonal terms of the one body density matrix, $D_{\alpha}$,
\begin{align}
\langle \hat{N}_{\alpha} \rangle = \sum_{i=1}^{N} D_{ii}^{\alpha}.
\label{DM}
\end{align}

\section{Results and Discussion \label{Results}}

\subsection{Molecules}
As a first test of our formalism, we apply our finite temperature \textit{ab initio} AFQMC algorithm to a range of molecular species, including first and second row atoms, water molecules, and the carbon dimer. These species were selected not only because of their ubiquity, but also because all but the carbon dimer are amenable to exact diagonalization in a minimal basis and can therefore be used for thorough benchmarking. In Table \ref{tab:Be}, we present the energies we obtain for the beryllium atom in the MIDI basis using exact diagonalization, our \textit{ab initio} finite temperature AFQMC algorithm, and mean field theory as a function of inverse temperature. 
\begin{table}[ht]
\centering
\begin{tabular}{C{1.2cm}|C{1.7cm}|C{2cm}|C{1.7cm}}
\hline \hline
1/$k_{B}$T & ED & AFQMC & MFT \\ \hline
0.01 & -10.81253 & -10.812(7) & -10.80377 \\
0.1 & -11.48668 & -11.48(2) & -11.39578 \\
1 & -13.99009 & -13.991(5) & -13.82935 \\
5 & -14.26004 & -14.26(2) & -14.17466 \\
10 & -14.39662	& -14.395(2) & -14.30915 \\
20 & -14.47835	& -14.482(2) & -14.43693 \\
50 & -14.48459	& -14.485(1) & -14.47476 \\
100 & -14.48460 & -14.485(2) & -14.47504 \\
\hline \hline
\end{tabular}
\caption{The internal energy of the Be atom at various inverse temperatures in the MIDI basis set, using ED (exact), AFQMC, and MFT. All energies are reported in Hartree.}
\label{tab:Be}
\end{table}
As discussed for completeness in the Supplement, our mean field theory replaces all two body operators with a product of a one body operator and a mean field. The reported exact diagonalization results have been produced by an in-house code and are what we use as our exact benchmark. All of the reported AFQMC results were obtained using $\Delta \tau=0.05$ for $\beta \ge 1$ (or $L=20$ for $\beta < 1$) with 128 walkers averaged over 20 blocks; any exceptions are noted in the captions. Please refer to the Supplement for details about the error analysis performed. As expected, mean field energies most align with ED results at high temperatures. In contrast, AFQMC energies are within milliHartrees of the exact results throughout the temperature range studied. Similar behavior is seen for He, Li, and H$_{2}$, as presented in the Supplement. We limit our comparison to ED to these species in this basis because of the relatively small number of orbitals involved; AFQMC can readily be applied to larger systems and its accuracy can be systematically improved by increasing the number of samples taken and decreasing the imaginary time step sizes used.  
\begin{figure}[H]
\includegraphics[width=0.5\textwidth]{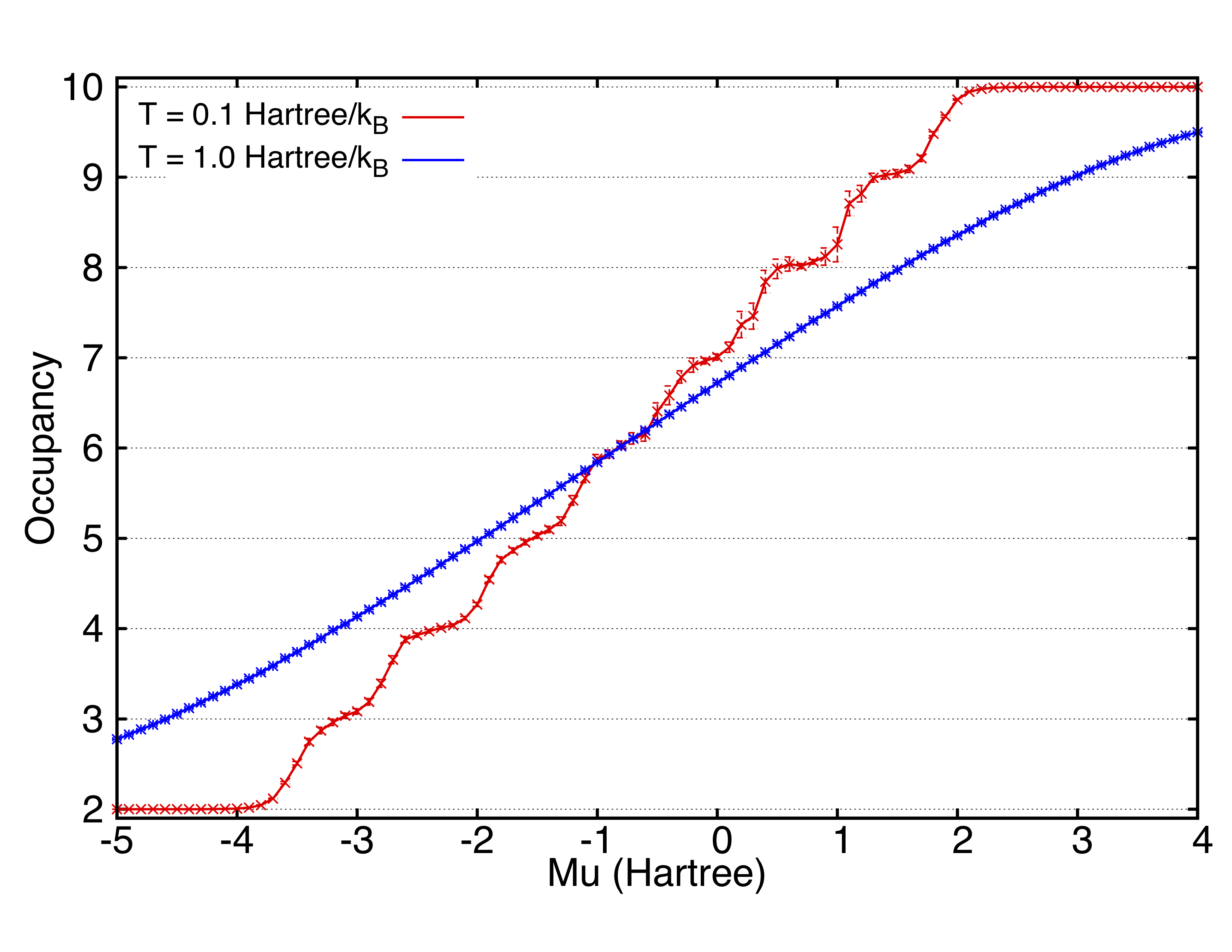}
\caption{Occupancy of the nitrogen atom at different chemical potentials for $T=0.1$ and $T=1.0$ Hartree$/k_B$, calculated using the STO-6G basis, averaged over 54 blocks. The occupancy is calculated by summing up the diagonal terms of the one body density matrix, as in Equation \eqref{DM}.}
\label{fig:occ_mu}
\end{figure}

Because this algorithm is constructed in the grand canonical ensemble, obtaining accurate energies for a given electron number necessitates determining the correct chemical potential to achieve that filling. In this work, we scan through chemical potentials to arrive at our desired occupancies. This is made possible by the emergence of a step-like profile reminiscent of the Fermi-Dirac distribution in occupancy vs. chemical potential plots at temperatures for which $k_BT \ll \Delta E$, where $\Delta E$ is the energy difference between states with different occupancies. This profile is depicted for the nitrogen atom in the STO-6G basis in Figure \ref{fig:occ_mu} at two different illustrative temperatures. At the lowest chemical potentials, all of the valence electrons in the nitrogen atom are stripped away, resulting in an average occupancy of two\footnote{Although not depicted here, if even lower chemical potentials were employed, even the two remaining core electrons would be stripped resulting in no occupancy whatsoever.} As the chemical potential is increased, the occupancy is 
increased in a continuous fashion at high temperatures and in an increasingly step-wise fashion at lower temperatures. These step-like plateaus make chemical potential searches not only feasible, but simple and highly accurate, circumventing the immediate need for a canonical ensemble formalism. 

\begin{figure}[ht]
\includegraphics[width=\textwidth]{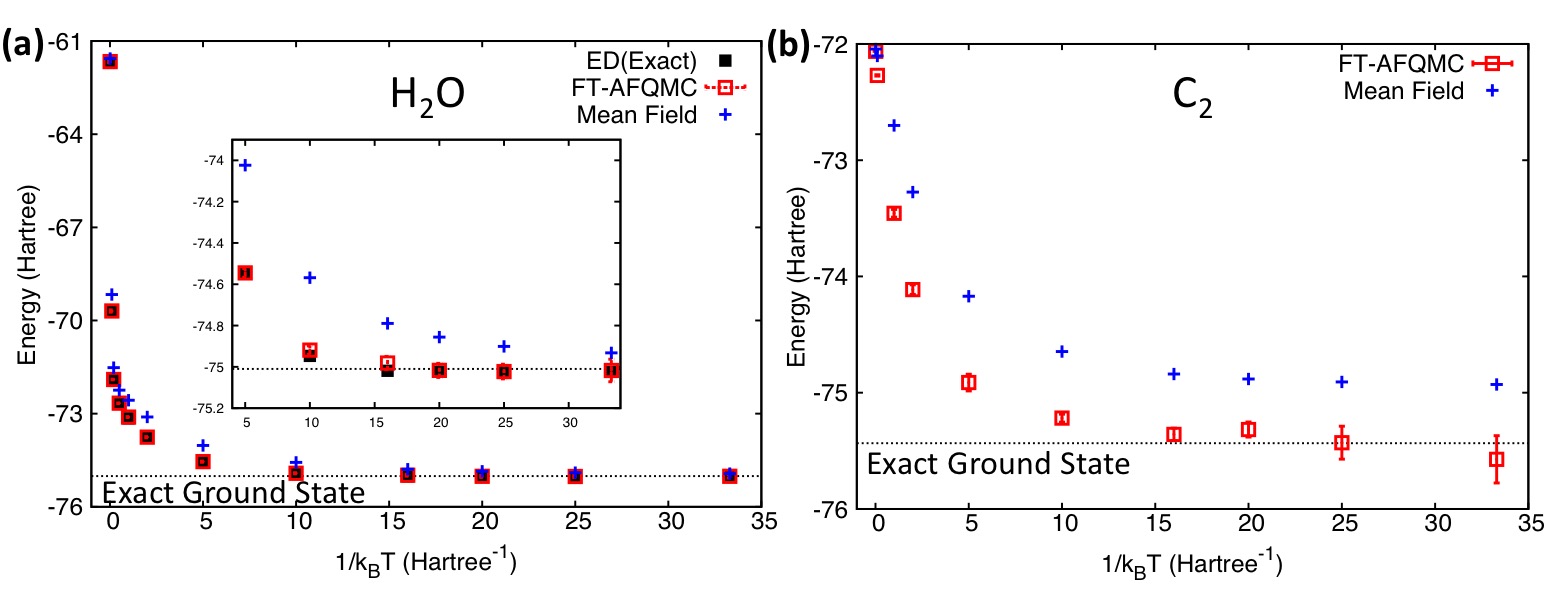}
\caption{Internal energy of the (a) H$_{2}$O (STO-3G basis) and (b) C$_2$ (STO-6G) molecules across a wide range of inverse temperatures using FT AFQMC, ED (exact), and mean field theory. The C-C and O-H bond lengths are 1.262 \text{\AA} and 0.96 \text{\AA} respectively, and the H-O-H bond angle is $109.5^{\circ}$. Exact ground state energies were obtained using the full configuration interaction routine in Molpro\cite{MOLPRO,MOLPRO-WIREs}; the full spectrum of exact excited state energies was too expensive to compute. The inset highlights the energies observed in the low temperature regime. AFQMC results for C$_2$ are averaged over 100 blocks. Note that the exact FT internal energies for C$_2$ cannot be obtained due to system size.}
\label{fig:H2O_C2}
\end{figure}

In Figure \ref{fig:H2O_C2}, we step beyond atomic species and present the internal energies of water and the carbon dimer as a function of inverse temperature. Overall, AFQMC agrees very well with the exact results across the whole temperature regime from high to low temperatures. In contrast, mean field theory tends to overestimate the internal energy, as it does not capture correlated contributions. Specifically, at high temperatures, these systems approach the classical limit and are dominated by one body kinetic and electron-nuclear contributions to the Hamiltonian, which are also well captured by mean field theory. As the temperature is lowered, the internal energy decreases as states with lower energies are favored. At the same time, the structure of the two body interactions becomes perceptible. In the low temperature limit, the system collapses almost entirely to the ground state and the energy approaches its zero temperature ground state value. The mean field energies also plateau at a value consistently larger than the exact result. Interestingly, as is evident in Figure \ref{fig:H2O_C2}, mean field theory makes the largest errors at intermediate temperatures. This is because, at intermediate temperatures, the system occupies a range of very specific excited states, a situation that mean field theory fails to capture. At the lowest temperatures, evidence of the phase problem begins to emerge, as manifested by increasingly large error bars. Even so, the results presented were obtained without the phaseless approximation. This is particularly fascinating because Table \ref{tab:Be} and Figure \ref{fig:H2O_C2} demonstrate that our finite temperature results come within milliHartrees of the exact ground state results. This suggests that finite temperature internal energies, which near the ground state energy before a pernicious phase problem crops up, may powerfully serve as a ``phase problem-light'' way of obtaining ground state energies without explicitly simulating the ground state. These plots may also hint at the fact that, in the ground state, averaging over blocks of simulations carried out to a finite $\beta$ before a catastrophic phase problem sets in may yield better statistics than carrying out continuous simulations to infinite $\beta$, as is commonly performed. It is moreover clear from Figure \ref{fig:H2O_C2} that C$_{2}$ is far more correlated than H$_{2}$O. Previous work suggests that C$_{2}$ contains a complex quadruple bond \cite{shaik2012quadruple}, which accounts for its more correlated behavior. Indeed, the difference between the first excited and ground state of H$_{2}$O (.3863 Hartree) is eight times greater than that for C$_{2}$ (.04539 Hartree). Regardless, AFQMC readily obtains highly accurate C$_{2}$ internal energies given sufficient averaging.  

\subsection{\textit{Ab Initio} Solids}

\begin{figure}[ht]
\includegraphics[width=\textwidth]{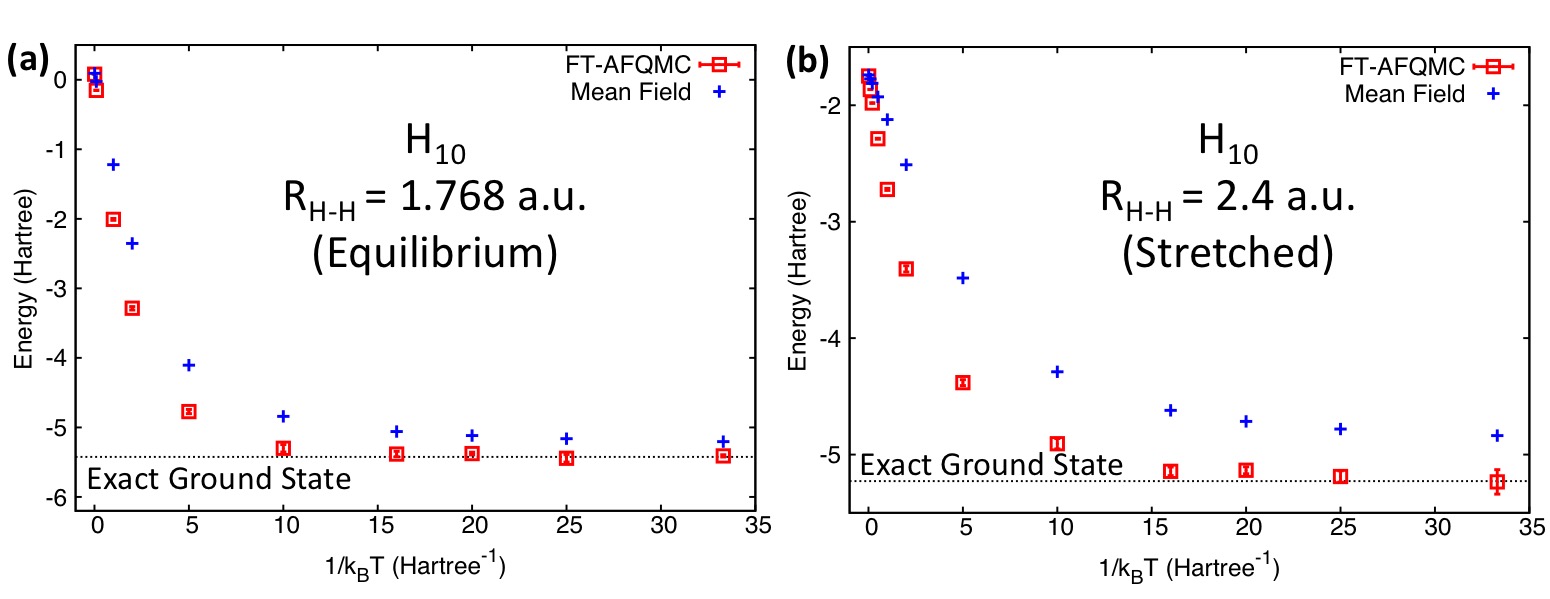}
\caption{Internal energy of the one-dimensional H$_{10}$ chain at its (a) equilibrium bond length and (b) an elongated bond length with the STO-6G basis set using FT AFQMC and mean field theory. The dashed black line labels the exact energy in the ground state \cite{motta2017towards}. Note that the exact FT internal energies cannot be obtained due to their prohibitive computational cost. AFQMC results for $k_B T<0.1$ Hartrees are averaged over 80 blocks.}
\label{fig:H10}
\end{figure}

As a demonstration of the applicability of our method to solids, we calculate the internal energy of a one dimensional H$_{10}$ chain at different bond lengths\footnote{Here, we constrain the geometry to have equal bond lengths between all hydrogens.} with open boundary conditions, as shown in Figure \ref{fig:H10}. The H$_{10}$ chain in a small basis may be viewed as the simplest representation of a solid that still retains non-density-density terms within its Hamiltonian \cite{motta2017ab}. Beyond its simplicity, what makes the hydrogen chain particularly illuminating is that its degree of correlation changes as its bond distances change. In particular, as the hydrogen chain evolves from the equilibrium geometry in Figure \ref{fig:H10} (a) to the stretched geometry in Figure \ref{fig:H10} (b), the overlap between adjacent hydrogen atoms decreases resulting in a smaller band gap and increased multi-reference character, which leads to a larger mean field error in Figure \ref{fig:H10} (b) than in Figure \ref{fig:H10} (a). Even so, FT-AFQMC is robust regardless of bond length, and by extension, metallic vs. insulating character. 

\subsection{Multi-Orbital Hubbard Model} \label{MHK}
Given the convincing results presented above for molecules and a model \textit{ab initio} solid, we also benchmark our algorithm on a notoriously strongly correlated model Hamiltonian, the two-band Hubbard-Kanamori Hamiltonian. The multi-orbital Hubbard-Kanamori model is a multi-band generalization of the Hubbard model that includes Hund's coupling terms pivotal to the accurate description of many transition metal oxides, such as the pnictides and ruthenates\cite{aoki2006triplet,koga2004orbital,piefke2011lda+,si2016high}. Its Hamiltonian may be expressed as
\begin{equation}
\hat{H}=\sum_{\langle i,j \rangle,\alpha,mm'}{t_{ij}^{mm'}\hat{c}_{im\alpha}^{\dagger}\hat{c}_{jm'\alpha}}-\mu \sum_{im\alpha}{\hat{n}_{im\alpha}}+\sum_{i}{\hat{h}_{i}}
\end{equation}
where
\begin{align}
\hat{h}_{i} &= U\sum_{m}{\hat{n}_{im\uparrow}\hat{n}_{im\downarrow}}+U'\sum_{m\neq m'}{\hat{n}_{im\uparrow}\hat{n}_{im'\downarrow}} 
+(U'-J)\sum_{m>m',\alpha}{\hat{n}_{im\alpha}\hat{n}_{im'\alpha}} \nonumber \\ 
 & +J\sum_{m\neq m'}{\big(\hat{c}_{im\uparrow}^{\dagger}\hat{c}_{im'\uparrow}\hat{c}_{im'\downarrow}^{\dagger}\hat{c}_{im\downarrow}+\hat{c}_{im\uparrow}^{\dagger}\hat{c}_{im'\uparrow}\hat{c}_{im\downarrow}^{\dagger}\hat{c}_{im'\downarrow}\big)}
\end{align}
is the local intra-site interaction for site $i$.
$\hat{c}_{im\alpha}^{\dagger}$ and $\hat{c}_{im\alpha}$ are the creation and annihilation operators for an electron with spin $\alpha$ at orbital $m$ of site $i$. $\hat{n}_{im\alpha}$ is the density operator of an electron with $\alpha$ spin at orbital $m$ of site $i$. $U$ is the on-site intra-orbital Coulomb repulsion, $U'$ is the on-site inter-orbital Coulomb repulsion, and $J$ is the Hund's exchange interaction. Assuming a spherically symmetric interaction and $t_{2g}$ wave functions (the typical choice), $U'=U-2J$\cite{georges2013strong}. While the $U$ and $U'$ density-density terms are readily amenable to standard DQMC, the Hund's exchange terms result in a substantial sign problem that thwarts the direct application of most QMC techniques\cite{Sakai_PRB_2004}. Nevertheless, these same terms may be viewed as particular instances of the full \textit{ab initio} Hamiltonian and consequently may be treated by the methods we have described above.   

\begin{figure}[ht]
\includegraphics[width=\textwidth]{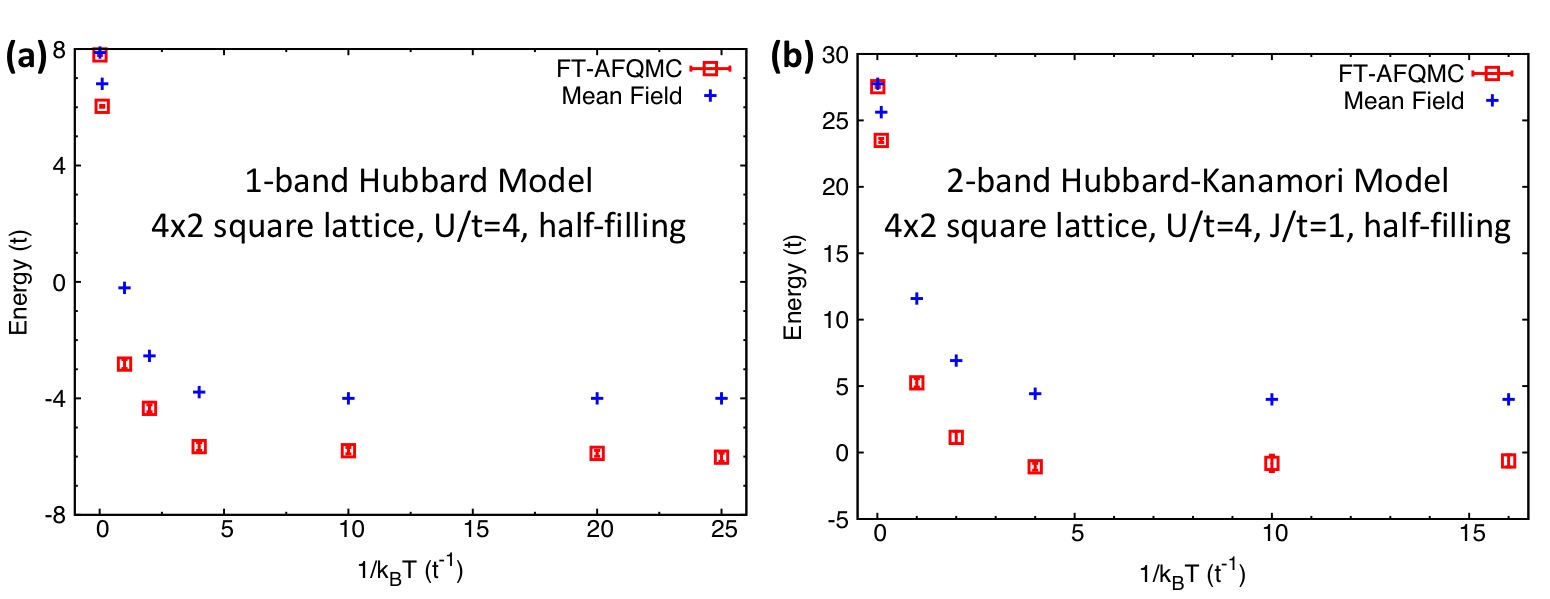}
\caption{The internal energy of (a) the $4\times2$, one-band Hubbard Model at U/t=4 and half-filling and (b) the two-band Hubbard-Kanamori Model at U/t=4, J/t=1, and half-filling averaged over 56 walkers for 100 blocks using variants of FT AFQMC and mean field theory over a range of temperatures. Exact FT internal energies were not obtained due to their exorbitant computational cost. The AFQMC results for the one-band Hubbard model are averaged over 10 blocks, while the two-band Hubbard-Kanamori model results are averaged over 100 blocks with 56 walkers.}
\label{fig:multiorbital}
\end{figure}

To demonstrate the applicability of our method to this more challenging model, we calculate its energy on a $4\times2$ rectangular lattice with intermediate Hund's coupling strength ($J=U/4$) at half-filling. Furthermore, we assume that hopping occurs only between adjacent sites within the same orbitals, i.e., $t_{ij}^{mm'} = t_{\langle i,j \rangle}\delta_{mm'}$. Comparing the top and bottom panels of Figure \ref{fig:multiorbital}, we see that the difference between the mean field and exact results for the two-band Hubbard Model is significantly larger than the same difference for the one-band Hubbard model, demonstrating that the two-band model is significantly more correlated than the one-band model. Despite having structurally similar two body terms in their Hamiltonians, the Hubbard-Kanamori model moreover manifests dramatically stronger correlation than the molecules illustrated above. The Hubbard-Kanamori model may therefore be viewed as a particularly challenging ``ab initio'' test case for algorithmic advances, such as improved trial density matrices, that may be needed for the more strongly correlated model systems to which we aspire to ultimately treat. Regardless, the convergence of the internal energies with decreasing temperature suggests that these models are within reach of our methodology.

\subsection{The Phase Problem}
As demonstrated in the previous sections, \textit{ab~initio} Hamiltonians generally result in a phase problem. To gain a quantitative understanding of the severity of the phase problem and therefore the efficacy of our method for the range of Hamiltonians studied, we analyze the weight distributions and average phase angles of our walkers for C$_2$ and H$_2$O using both the free propagation and background subtraction versions of our algorithm, as well as for H$_{10}$ at equilibrium and stretched bond lengths using just the background subtraction version of our algorithm. We do not present Hubbard or Hubbard-Kanamori phase angles here because these models do not possess a phase problem for the fillings employed in Section \ref{MHK}; we save a discussion of non-half-filling Hubbard-Kanamori results that possess a phase problem for a subsequent paper. 

\begin{figure}[ht]
\includegraphics[width=\textwidth]{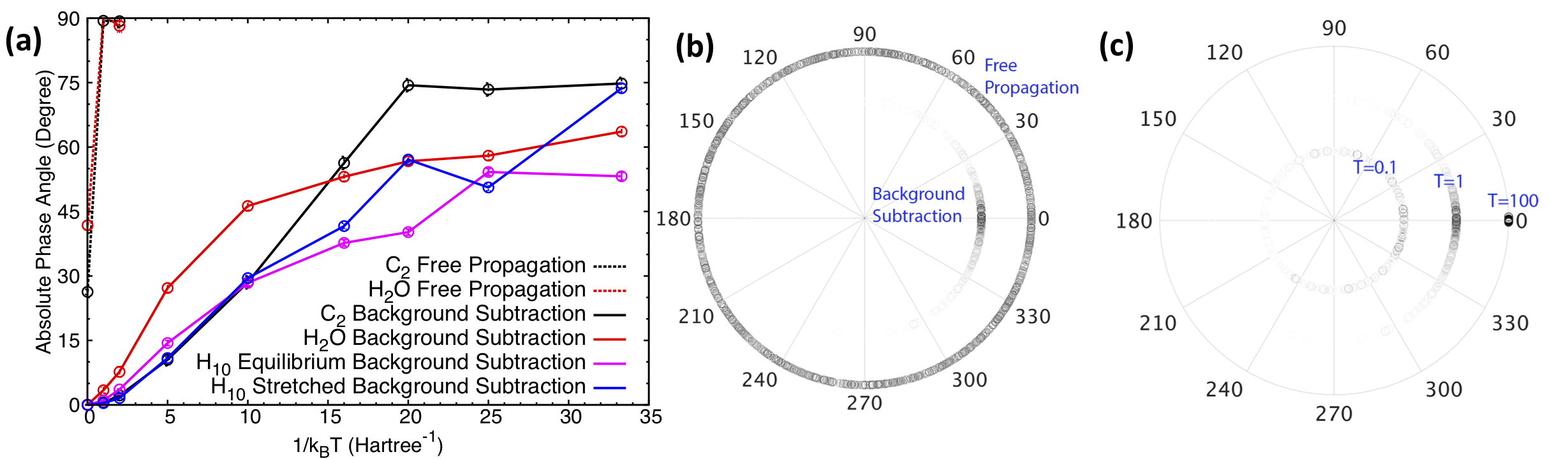}
\caption{(a) The average absolute value of the phase angle at different temperatures for the various systems investigated. Dashed lines represent the free propagation results, while solid lines denote background subtraction results. (b) The walker weight distribution of free propagation (outer circle) and background subtraction (inner circle) samples in polar coordinates for H$_2$O at $T=1$ Hartree/$k_B$.(c) The walker weight distribution in polar coordinates for H$_2$O at different temperatures (outer circle for high temperature, and inner circle for lower temperature) with background subtraction. In both (b) and (c), each empty circle represents one walker. Each circle's polar angle denotes its walker's complex phase, while each circle's shading scales with the absolute value of its walker's weight. 1280 walkers are plotted in each case.}
\label{fig:phase}
\end{figure}
As is illustrated in Figure \ref{fig:phase}(a), free propagation (dashed lines) leads to phase angles that rapidly increase with inverse temperature, reaching $90^{\circ}$ even after just a few inverse Hartree. In contrast, when background subtraction (solid lines) is applied, the phase angle increases at a much slower pace, so slow that even at low temperatures close to ground state, the phase problem is still tractable. Interestingly, the phase problem appears to plateau in these systems at intermediate inverse temperatures, suggesting that most of the phase problem stems from propagating down from a larger ensemble of states to a small subset of excited states, not from resolving between a handful of states and the ground state. This is consistent with our previous observation that the largest amount of correlation is present at the same intermediate inverse temperatures. To better understand the effect of background subtraction as compared to free propagation, Figure \ref{fig:phase}(b) plots the walker weight distribution employing both techniques for H$_2$O at $T=1$ Hartree/$k_B$. When freely propagated, walker weights tend to distribute evenly among all phase angle values (the outer circle) and they largely cancel each other in Monte Carlo averages, resulting in a very severe phase problem. Once background subtraction is applied (inner circle), all the walker weights tend to cluster around a phase angle of zero, with walkers with larger absolute weights carrying smaller absolute phase angles. This greatly reduces the cancellation in Monte Carlo averages, and improves the sampling efficiency. Lastly, Figure \ref{fig:phase}(c) demonstrates the severity of the phase problem at different temperatures taking H$_2$O as an example. In the high temperature limit, nearly all the walker weights are distributed on the real axis with similar absolute weights, as shown by the overlapping dark black circles on the outermost ring at 0 phase angle for $T=100$ Hartree$/k_B$. As the temperature is lowered to $T=1$ Hartree$/k_B$ (middle ring), the walkers develop larger phase angles and spread over a larger portion of phase angle space. When the temperature is decreased further to $T=0.1$ Hartree$/k_B$ (inner ring), walker weights nearly uniformly populate the full complex plane. Altogether, the fact that the phase angle is well-behaved for the disparate \textit{ab initio} examples studied here bodes well for this method. 

\section{Conclusions \label{Conclusions}}
In summary, we report an \textit{ab initio} FT-AFQMC method capable of high accuracy predictions across a wide range of temperatures for molecules, solids, and model Hamiltonians alike. We find that for all of the systems presented, the phase problem at finite temperature is significantly smaller than the corresponding ground state phase problem and can therefore be effectively controlled not only without invoking the phaseless approximation, but also only making use of easy-to-acquire mean field trial density matrices. This suggests a path toward exploiting finite temperature (or fixed $\beta$) simulations to acquire ground state information. Our algorithm moreover benefits from the low $O(N^{3})-O(N^{4})$ scaling characteristic of other AFQMC algorithms, which places warm dense matter, solid state, and condensed matter applications within reach. We look forward to \textit{ab initio} FT-AFQMC explorations of these systems in the near future. 

\begin{acknowledgement}
B.R. and Y.L. thank James Shepherd, Garnet Chan, Shiwei Zhang, and Hao Shi for stimulating discussions, and Tingwei Meng for providing a proof for the positive semi-definite property of the \textit{ab initio} supermatrix. Y.L. thanks the Brown Presidential Fellowship and B.R. thanks NSF grant DMR-1726213 for financial support. This research was conducted using computational
resources and services at the Brown University Center for Computation and Visualization.
\end{acknowledgement}

\begin{suppinfo}
Additional theoretical details and tabulated data.
\end{suppinfo}

\bibstyle{achemso}
\bibliography{abinitio}

\begin{figure}[ht]
\includegraphics[width=\textwidth]{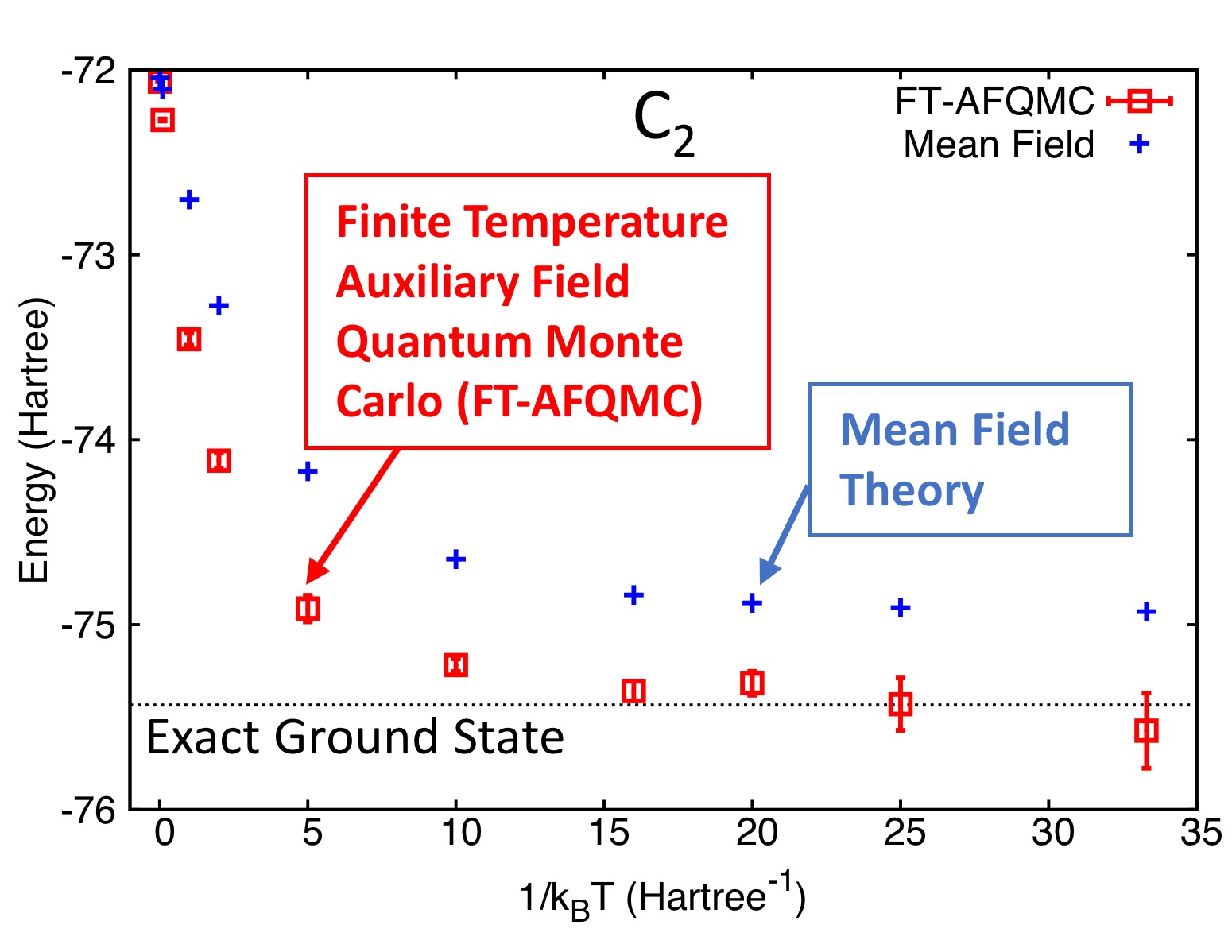}
\caption{Table of Contents (TOC)}
\end{figure}

\end{document}